\documentclass[10pt]{article}
\usepackage[british]{babel}
\usepackage{amssymb}
\usepackage{amsmath}
\usepackage[dvips]{graphicx}
\usepackage{epsfig,bm}
\usepackage{vmargin}
\usepackage{pdfsync}
\usepackage{enumerate}
\usepackage{amsfonts}
\usepackage{graphics}
\usepackage{color}
\usepackage{setspace}
\usepackage{amscd}
\usepackage{pstricks}
\usepackage{lscape}
\usepackage{enumitem}
\usepackage{amsxtra}
\usepackage{amstext}
\usepackage{latexsym}
\usepackage{verbatim}
\usepackage{dcolumn}
\usepackage{bm}
\usepackage[numbers,sort&compress]{natbib}
\usepackage{tabularx}
\usepackage{graphics}
\usepackage{subcaption}
\usepackage{array}
\usepackage{amssymb}
\usepackage{graphicx}
\usepackage{caption}
\usepackage{lineno}
\date{}
\title{Markovian language model of the DNA and its information content}
\author{S. Srivastava$^{1}$ and M. S. Baptista$^{1}$}

\begin{document}
\maketitle
\begin{center}
$^{1}$ Institute for Complex Systems and Mathematical Biology, SUPA, University of Aberdeen, Aberdeen, AB24 3UE, United Kingdom
\end{center}
\begin{abstract}
This work proposes a markovian memoryless model for the DNA that simplifies enormously the complexity of it. We encode nucleotide sequences into symbolic sequences, called words, from which we establish meaningful length of words and group of words that share symbolic similarities. Interpreting a node to represent a group of similar words and edges to represent their functional connectivity allows us to construct a network of the grammatical rules governing the appearance of group of words in the DNA. Our model allows  to predict the transition between group of words in the DNA with unprecedented accuracy, and to easily calculate many informational quantities to better characterize the DNA. In addition, we reduce the DNA of known bacteria to a network of only tens of nodes, show how our model can be used to detect similar (or dissimilar) genes in different organisms, and which sequences of symbols are responsible for the most of the information content of the DNA. Therefore, the DNA can indeed be treated as a language, a markovian language, where a "word" is an element of a group, and its grammar represents the rules behind the probability of transitions between any two groups. 
\end{abstract}

\section{Introduction}\label{section_introduction}
One of the most studied complex systems in biology is the genome of living organisms, composed by Deoxyribonucleic Nucleic Acid (DNA). The central dogma of life is related to how DNA transcribes into mRNA which finally translates into proteins. A big challenge is to understand the complexity of the dynamical organisation of the DNA, a system that is the result of millions of years of evolution. The genome of any organism is its hereditary information. It is encoded either in the form of DNA composed of four different types of chemical molecules  $[$\textit{Adenine (A), Guanine (G), Cytosine (C)and Thymine(T)}$]$ or in the form of Ribonucleic Nucleic Acid (RNA) as present in many viruses. The genome includes both the genes and the non-coding sequences of the DNA/RNA \cite{nelson2010lehninger}. Sequencing and high-throughput experiments have contributed much to the genomic data in the last 10-15 years. Still the data has to be analysed \cite{Kim:2007}.

Natural language analysis has been a topic of interest in the last decade \cite{Berger1996,Lewis96naturallanguage,Harispe2015}. Natural language written texts can be considered as being composed by a series of letters, syllables, words or phrases. During the 19$^{th}$ century many linguists like Schleicher and Haeckel interpreted language as a living system \cite{Richards1989}. Based on this concept Darwin also proposed that evolution of species and language are similar \cite{Darwin1916}. Many researchers have introduced the concept of linguistic into biology \cite{Botstein1997}. Brendel et al. used formal linguistic concepts to define a basic grammar for genes, based on the idea that mutating a piece of genetic information was similar to modifying words \cite{Brendel1986}. Similar to works that aimed at finding the relevant words, their relationships and their information content in natural languages, many studies have focused on analysing genomic sequences like DNA and proteins as if they were a language, using similar methodological approaches as the one used to model natural languages \cite{Botstein1997}. Formal linguistic concepts were used in Ref. \cite{Brendel1986} to define basic grammatical rules that describe how genes can mutate, inspired in the grammatical rules of languages that regulate how and which word follows a previous word. Gramatikoff \textit{et. al} \cite{Gramatikoff} have used lexical statistics to identify and represent structural, functional and evolutionary relationships for multiple genomic sequences and texts from natural languages. 

DNA sequences have also been analysed using approaches to characterise complex systems, for example by converting the DNA sequences to numerical signals using different mappings \cite{Vera2009,Anastassiou2001,Buldyrev1995,Cristea2003}. A commonly used mapping is to convert the DNA into binary sequences \cite{Voss1992}. Other mappings of the DNA look for spatial patterns by considering inter-nucleotide distances \cite{Nair2005,Vera2009,Baptista2000}. These models of the DNA capture the recurrence property of codons (a word formed by 3 nucleotides) \cite{Baptista2000} by measuring the statistics of the symbolic distance separation between two codons. In the work of Ref. \cite{Hao2007}, they have shown how to analyse the long DNA sequences by converting it to an image. The DNA was characterized by the fractal-like patterns appearing in this image as a result of the forbidden words.

Our model is constructed using some concepts and tools from Ergodic Theory and Information Theory to interpret genomic data (nucleotide sequences) as if it were a language that can be analysed by the tools of symbolic dynamics. Each language has its rules. Our motivation in this work is to propose and study a meaningful language for the DNA. To establish a language for the DNA, we specify the length of words and a set of relevant groups of words, and create a network of words from functional connections, linking how topological complexity in the functional networks arises and how this complexity is connected to the complexity of life (production of proteins). In order to achieve this goal, we analysed the genome of \textit{Escherichia coli} or \textit{E.coli}, \textit{Shigella dysenteriae}, \textit{Rhodococcus fascians} and \textit{Saccharomyces cerevisiae}. These organisms are commonly used model systems: their genome and genes are well known, well studied, and can be used to test mathematical approaches towards modelling the DNA. Firstly, we represent the genome of these organisms on a symbolic space. The words of the genome are encoded in such a way that the nucleotide sequences are represented by real numbers that can be plotted in a symbolic space. This space allows a straightforward characterisation of the DNA through informational quantities (Shannon entropy rate, mutual information rate, and statistical measures), and ergodic quantities (correlation decay, and transition probabilities). To group the words and specify their lengths, we find a partition of the symbolic space composed by $N^2$ equal boxes. A box is a region in this symbolic space whose points within encode and define a group of words of length $2L$ that are all formed by the same small sequence of symbols with length $2L_{n}$, where $L_n = \frac{1}{2}\log_{2}(N)$, $N=4^i$, $i \in N$, and $L_{n}<L$. To create a grammatical network of words of the DNA, we set the nodes to represent groups of words and the edges to represent the grammatical rules governing the transition probabilities between group of words.  

Our Markovian model of the DNA is constructed by searching for the existence of a Markov partition (See Sec 3(b) and Supplementary Material) in the symbolic space for which the correlations of the points within boxes of this partition separated ``in time'' decays to approximatively zero, where ``time'' denotes the nucleotide distance ($2L_{n}$) between 2 words encoded by two points in distinct boxes. This finding allow us to see the DNA as an approximate Markov process in which the behaviour of the whole system can be approximately described by the transitions among the group of words of finite-length. It was shown however by Arnedo \textit{et. al} \cite{Arneodo2011} that long words composed of more than 100 bp in a DNA separated by many nucleotides have slow power-law correlation decay. Complementing the results in Ref. \cite{Arneodo2011}, Buldyrev et. al. in Ref. \cite{Buldyrev1995} have analysed the DNA sequences (words) and have shown that there is no correlation between sequences of  nucleotides with length between 10 to 100 bp. Our Markov model for the DNA has memoryless properties. It is however based in the relationship between groups of words, not between words as in Refs.\cite{Arneodo2011} and \cite{Buldyrev1995}. The correct choice for grouping the words provide the memoryless property of our model. It is however based on the relationship between groups of words, not between words. The transformation of a system that has correlation into a memoryless symbolic system is a normal procedure to study chaotic systems by using a symbolic representation of its trajectory. The correlation between time separated points in chaotic systems or correlation between symbols created by a non Markovian partition of the phase space decays to zero certainly after an infinitely long time \cite{Anishchenko2003,Feldman2012}. However, the symbolic sequences generated by Markov partitions of chaotic systems have sequences whose correlations decay to zero even for a  finite and small time separation, and for words of finite and small length. 

Then, along the lines of the work in Ref. \cite{Sinatra2010}, we apply our model to create a network representation of the DNA, where the nodes represent the different group of words and edges represent the probability of transition between two group of words that are strongly correlated and that are separated by 1 nucleotide. Preserving the connections responsible for most of the information of the DNA, we create reduced network models of the DNA, which capture the most relevant features of it, and is able to identify hub groups of words, which have a similar function to the core words identified in Ref. \cite{Gerlach2013} in natural languages. This network model of the DNA allows us to create a similarity measure to identify genes in various genomes. We analysed genomes of \textit{E. coli}, \textit{S. dysenteriae}, \textit{R. fascians} and \textit{S. cerevisiae}. We observed that \textit{E. coli} and \textit{S. dysenteriae} are very similar to each other and there were remarkable differences between \textit{E. coli}, \textit{S. cerevisiae}. 

In our Markov model, we use finite-length words which have all the same length, in contrast to the work in Ref. \cite{Kalimeri2012} that considers variable word length. Our model allows for an easy calculation of entropy rates from group of words with short length, a quantity that is usually calculated over very long words demanding high computational costs. In our model of DNA, there are forbidden words as in Ref. \cite{Hao2007} but no forbidden group of words, all words contain and are formed by the same small symbolic sequence length of $\log_{2}(N)$, and any two words separated by $\log_{2}(N)$ nucleotides are roughly decorrelated. There are $N^2$ group of words.

This paper is organised as follows. Our Markov model of the DNA is presented in Sec. 3. In Sec. 3(a), we show how to define a word, and in Sec. 3(b), we show how to group them. The network description of the DNA with a comprehensive analysis of the words that are mostly correlated in presented in Sec. 4, and the validation of the model and a discussion about the relationship between words in our model and the biological function behind them is presented in Sec. 5. 

\section{Symbolic representation of the DNA and a partition}\label{section_symbolic_representation_of_the_DNA}
Symbolic dynamics is a way to represent an orbit of a dynamical system whose points belong to the real set by a sequence of symbols taken from a finite set. Assume that a dynamical system is represented by (${X,T}$), where ${X}\in\mathbb{R}^{d}$ is a set and ${T}$ is a transformation that operates on X. \textit{T} maps \textit{X} to itself. The set \textit{X} is the set of all possible states of a system and the transformation \textit{T} evolves the state of the system, \textit{X}, in time \cite{Brian:1995,Feldman2012}. The first step into encoding an orbit by a symbolic sequence is the specification of a partition. This can be done by dividing ${X}$ into cells, where a point within a cell would represent a symbolic sequence. The main motivation for using symbolic dynamics comes from the idea of observing an infinite resolution orbit by making finite resolution observations (represented by a finite alphabet of symbols) and being able to describe the properties of the system being observed \cite{Crutchfield}. This symbolic sequence is then encoded into a sequence of numbers, such that a symbolic space can be constructed. In this space, the transformation representing how points are mapped preserves the main features of the transformation \textit{T} of the original space. 

For our problem of modelling the DNA, the symbols are given. Our main interest is to define a biologically relevant encoding that creates a symbolic space whose distance between points is a measure of similarity between two sequences of 2L symbols, here denoted as words. In order to have a 2D symbolic space, we construct two symbolic sequences, called the \textbf{past} and the \textbf{future} symbolic sequences. We first represent the nucleotides by natural numbers: (A,T,G,C)=(0,1,2,3). Given a symbolic sequence in the DNA with a length of 2L and assuming L=3, if the DNA sequence is AATCGT, then the past sequence is AAT and the future sequence is CGT. The integer representation of this DNA sequence is given by (\textit{$s_{0}$}, \textit{$s_{1}$}, \textit{$s_{2}$}. \textit{$s_{3}$}, \textit{$s_{4}$}, \textit{$s_{5}$}) = (001321). The encoding into real numbers of a symbolic sequence is made as in the following assuming we are at the position $j^{th}$ of the DNA (the location of the $j^{th}$ nucleotide) then, the past symbolic sequence is encoded by

\begin{equation} \label{eq:past}
\delta_j  =\sum_{i=1}^{L}\frac{s_{j-i} \ 4^{L-i}}{4^{L}-1},
\end{equation} 

\noindent
and the future sequence is encoded by
\begin{equation} \label{eq:future}
\gamma_j  =\sum_{i=1}^{L}\frac{s_{j+i-1} \ 4^{L-i}} {4^{L}-1}, 
\end{equation} 
\noindent
where $4$ is the number of symbols encoding each nucleotide and $j \geq L$. A $2L$ length word centred at the position $j$ is encoded by the  point ($\delta_{j},\gamma_{j}$). This word is $\tau$ nucleotides apart from a word centred at $j+\tau$ that is encoded by the point  ($\delta_{j+\tau},\gamma_{j+\tau}$). We adopt a dynamic description for our symbolic space, meaning that we assume that the point ($\delta_{j},\gamma_{j}$) is mapped to the point ($\delta_{j+\tau},\gamma_{j+\tau}$) after $\tau$ ``time'' iterations. Both $\delta$ and $\gamma$  $\in$ ${\rm I\!R}$ [0,1]. The encoding proposed in Eqs \eqref{eq:past} and \eqref{eq:future} is a standard encoding of a quaternary alphabet into a set of real numbers. The encoding of symbolic sequences into real numbers needs to satisfy some conditions in order to capture the underlying deterministic behaviour of the DNA: the symbolic space is independent on the starting point (so we can study equivalent spaces for the genes and the whole DNA or genome); the encoding produces similar results for palindromic (mirror images), which means that correlation decays as you go further into the past or further into the future; the encoding has finite-length words, creating a symbolic space whose distance between points in one coordinate is no smaller than 4{$^{-L}$} (or not smaller than 2{$^{-L}\sqrt{2}$} in the 2D space); one-to-one coding, such that a symbolic sequence will be uniquely encoded into a real number.

\begin{figure}[!htb]
\begin{subfigure}{0.3\textwidth}
\caption{}\includegraphics[scale=0.3]{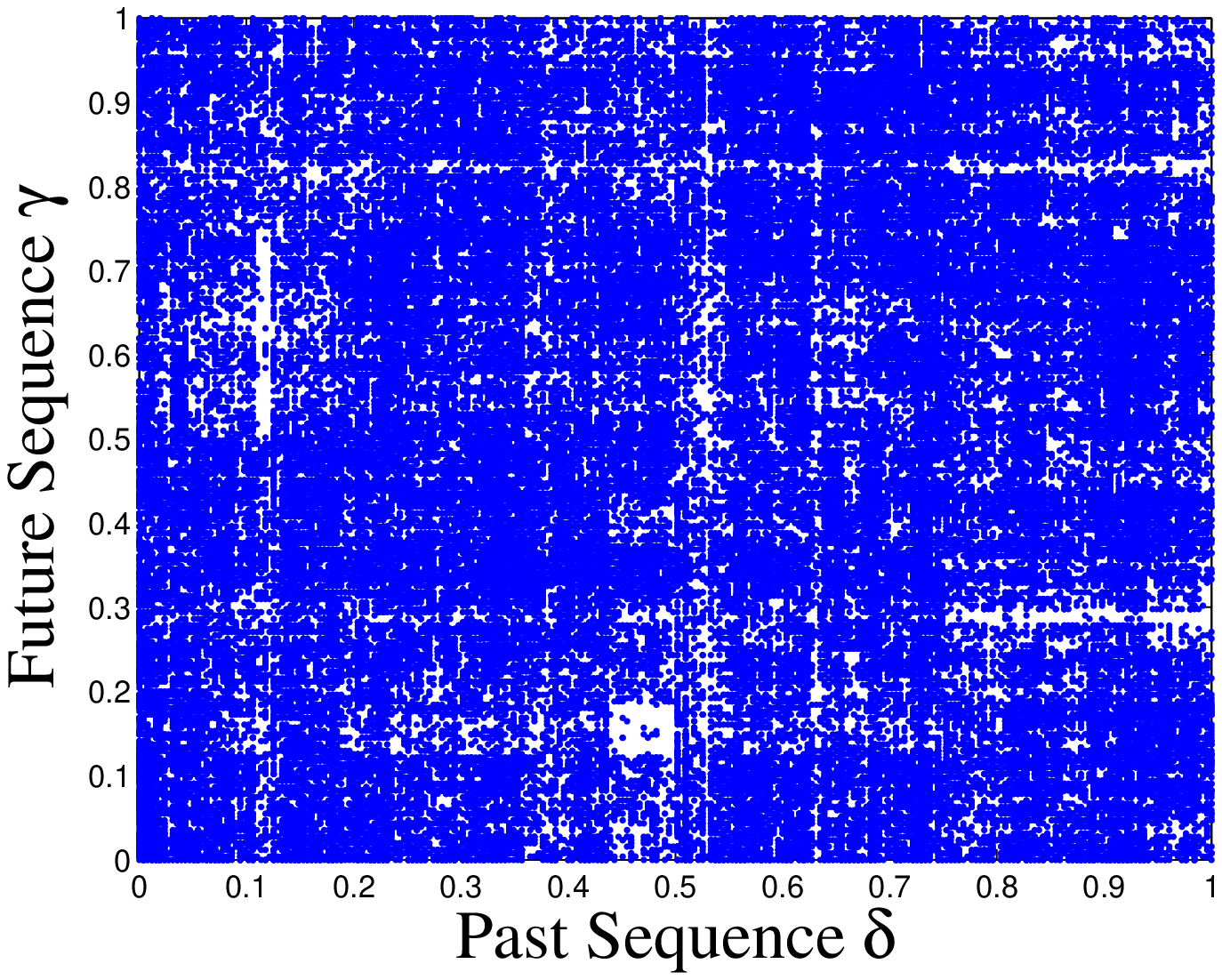}
\label{fig:phase_space}
\end{subfigure}
\hspace*{\fill}
\begin{subfigure}{0.3\textwidth}
\caption{}\includegraphics[scale=0.3]{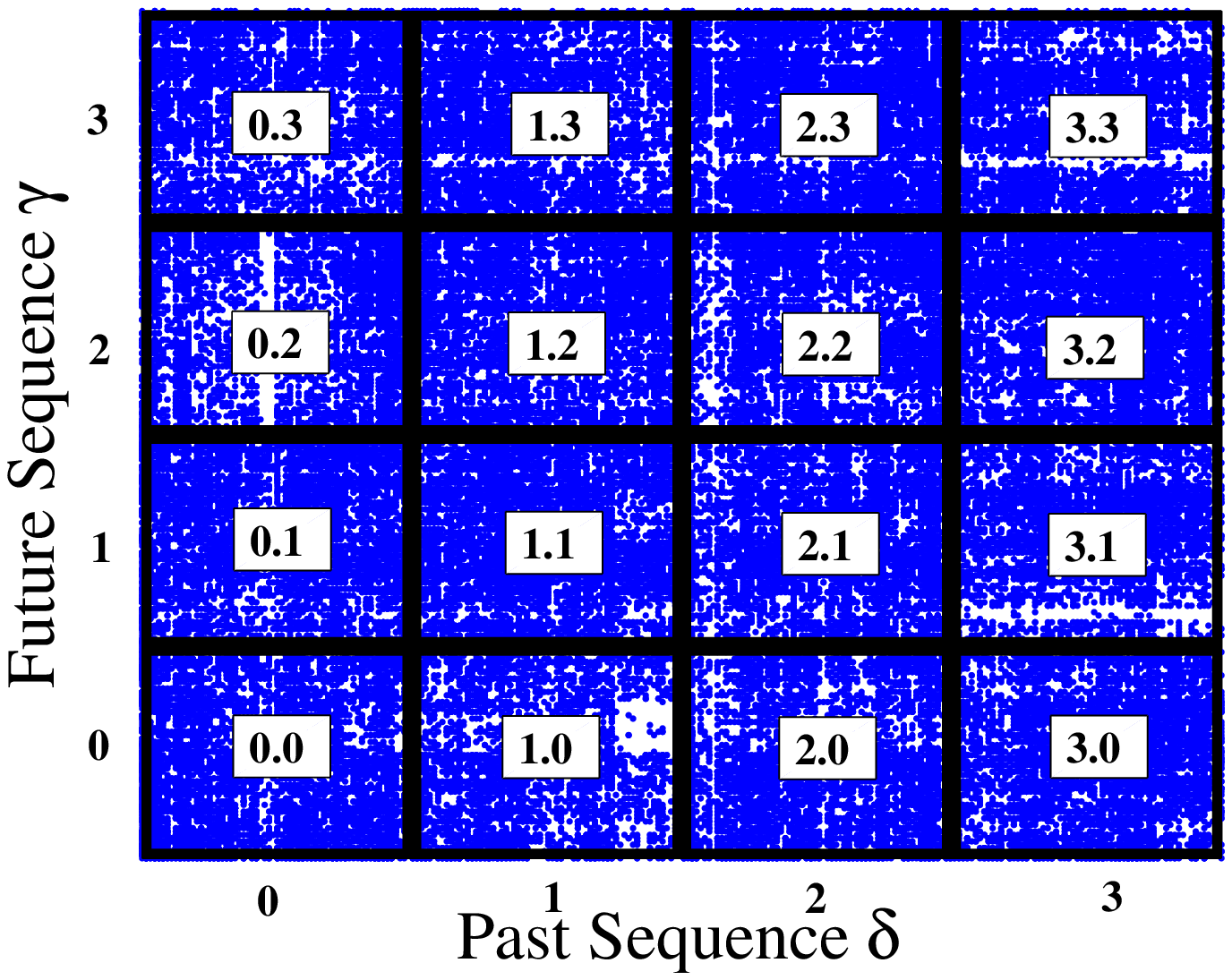}
\label{fig:word_symbolic_space}
\end{subfigure}
\hspace*{\fill}
\begin{subfigure}{0.3\textwidth}
\caption{}\includegraphics[scale=0.3]{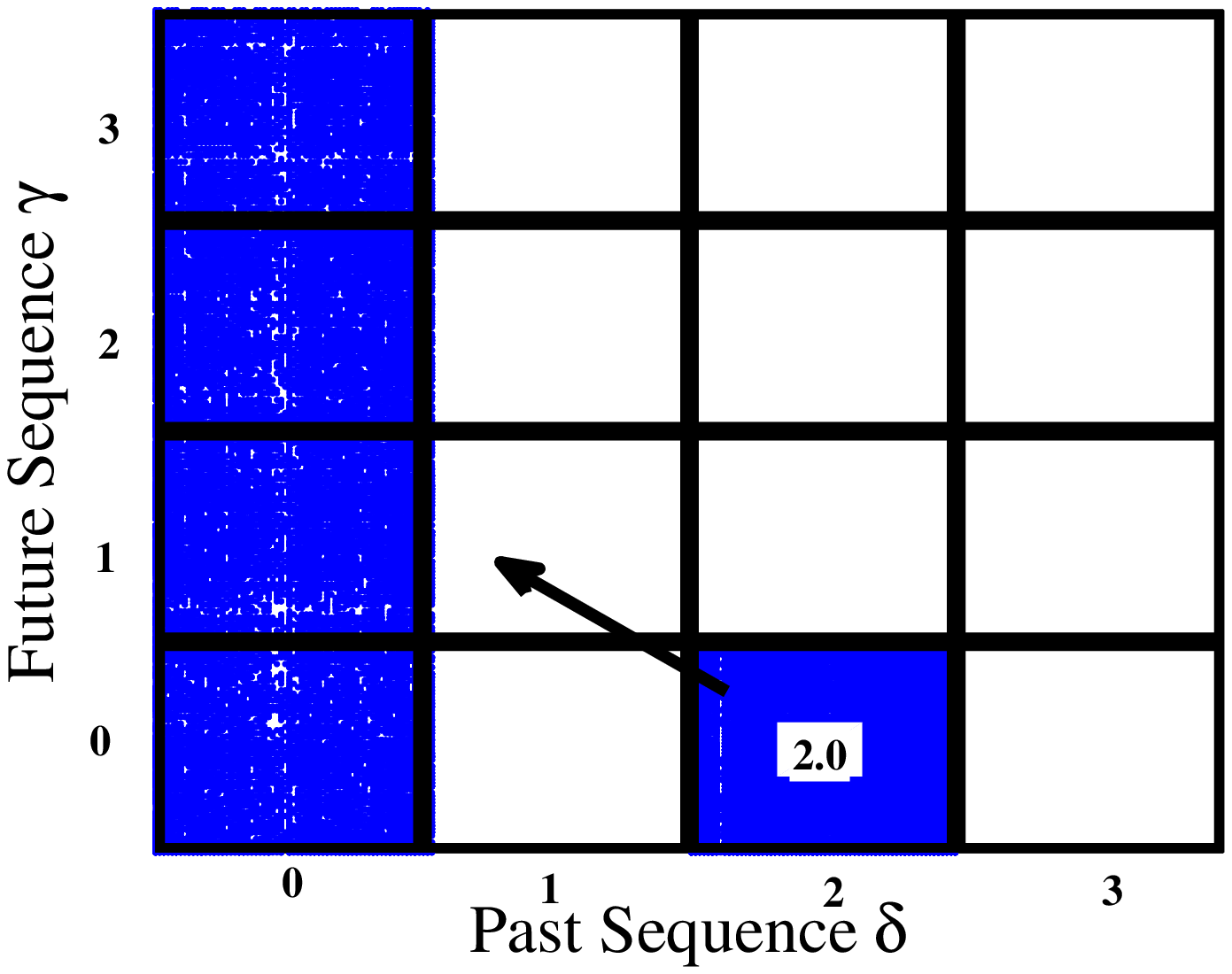}
\label{fig:transition}
\end{subfigure}
\caption{Symbolic space of the DNA of $E.coli$ with horizontal axis representing the encoding of past sequences $\delta$ and vertical axis representing the encoding of future sequences $\gamma$. The grid lines represent a partition with $N$=4. (a) Symbolic space of the DNA of $E. coli$ where existing words are represented by gray points (blue online) and white space represents the presence of forbidden words, (b) Symbolic space with a grid partition where the boxes define regions of points that encode similar words forming a group, (c) Symbolic space representing how points from a box move to 4 other boxes.} \label{fig:sym_space}
\end{figure}

A 2D picture (Fig. 1(a)) showing all the point coordinates ($\delta$,$\gamma$) is called \textbf{symbolic space of the DNA}. Figure \ref{fig:sym_space} horizontal axis corresponds to the encoding of the past sequences, $\delta $, and vertical axis corresponds to the encoding of the future sequences, $\gamma$. We notice some very dense areas in this space which means that many similar nucleotides sequences exist in the genome. There are some blank spaces, which represent forbidden sequences in the DNA. In Fig. 1 (b), we place a grid of $N^2$ boxes ($N$=4) into the symbolic space. There are $N$ columns and $N$ rows. A column in this space represents boxes where points encode symbolic sequences that contain the same length $L_{n}$ past symbolic sequence. A row represents all the boxes whose points encode symbolic sequences that contain the same length $L_{n}$ future symbolic sequence. Each box contains points that encode a group of words that are similar, i.e. they all have the same length $L_{n}$ past and future symbolic sequence. Let us understand the properties of symbolic trajectories in the symbolic space. In the partition of Fig. 1(b), a point within the box with coordinates $\delta\in$[0, 0.25] and $\gamma\in$ [0, 0.25] belongs to the box named `0.0'. Points within are mapped after 1 iterations (or 1 shift in the symbolic sequence) to the partition `0.X' where X $\in$ [0,1,2,3]. After 2 iterations they can be mapped everywhere in the symbolic space. Notice that a point inside box `2.0' represents a symbolic sequence of length $2L$ in the DNA ($L_n < L$), where first past symbol is `2' and first future symbol is `0'. This sequence is centred at a location $i$ of the DNA, has at the location $i-1$ a nucleotide `G' and at location $i$ the nucleotide `A'. Every $2L$ length symbolic sequence encoded by a point that belongs to the box ``2.0'' belongs to the group of words `2.0', having a $L_{n}=1$ past symbol `2' and a $L_{n}=1$ future symbol `0'.  As illustrated in Fig. 1(c), points from box `2.0' iterate to `0.X', where `X' can be either 0, 1, 2 or 3. The length of the symbolic name of a box in a partition is given by $2L_n$, where $L_n < L$.

\section{Determination of the approximate Markov partition: specification of word length and grouping}\label{section_Markov_length}

\subsection{Determination of word length}\label{subsec:determination_word_length}
The symbolic space allow us to find the hidden patterns of the DNA, since two similar symbolic sequences should be encoded by two nearby points in the symbolic space. We initially consider the DNA of $E. coli$. To have an estimation of the length of the words ($2L$) that our model should consider, we notice that the genome size of ${E.coli}$ is 4.6 million bp which means that it has ${4,600,000}$ symbols. According to the encoding rules of Eqs. (\ref{eq:past}) and (\ref{eq:future}), the  maximum number of different sequences of length $L$ would be ${4,600,000-2}L$. But for a given $L$, the total number of combinations possible is ${4{^{2L}}}$ sequences. Therefore, ${4{^{2L}}}$ should be at least smaller than ${4,600,000}$ if we were to have each sequence to appear once. Then, ${4{^{2L}}< 4600000}$ and therefore ${2L<11.06}$. Our main interest is to search for invariant properties in the language of the DNA, in order to obtain a stationary Markov model of it \cite{Rubino2014}.  In order to achieve that, we determine the appropriate value of $L=L^{*}$  by the largest length that allows the topological entropy rate of the symbolic space to remain invariant as $L$ is changed. If the length of the sequence is larger than $L^{*}$ then the symbolic space properties change abruptly. Its statistical properties would no longer be invariant by a change in the length. As shown in the section `Determination of word length' in Supplementary Material, the appropriate $L^*$ to be used in this work is $L=L^*=4$.

In contrast to the work of Ref. \cite{Kalimeri2012}, we consider finite words with the same length, which allows for an easy calculation of entropy rates from group of words of small short length, a quantity that is usually calculated over very long words demanding high computational costs.

\subsection{Determination of grouping of words and the order of partition}\label{determination_of grouping}
A group is a set of symbolic sequences whose encoding ($\delta$,$\gamma$) points fall within the same box of the partition. Let us now determine the optimal partition. In the symbolic spaces of the DNA, the minimum distance of points is equal to $4^{-L}$, i.e., the minimum distance between points in a horizontal or vertical direction. We partition the symbolic space in ${N^2}$ equal boxes with sides of $\epsilon=\frac{1}{N}$.

To estimate $N$, we require that $\epsilon\gg4^{-L}$, so $N\ll4^L$. An orbit in the symbolic space is constructed by a series of shift operations (from left to right) in the symbolic sequence. Given points in a box, an order-$T$ partition is such that after $T$ shifts in the symbolic sequence (or after $T$ iterations of points in the symbolic space) these points spread out to the whole symbolic space. If a partition with $N^{2}$ boxes is well chosen, the correlation between the points of a box (encoding words) and the points of the boxes containing $T$ forward iterations of these points (or words separated by a distance of $T$ nucleotides) decays to approximately zero for a small finite ``time'' $T$. For an order-$T$ Markov partition, the correlation between initial points, and their $T$ forward iterations decays to zero for the finite time $T$. Our goal is to construct an approximation of a Markov partition model to the DNA \cite{Bolt2005, Bolt2008}. Since the dynamics generating the symbols are not known, our model can also be considered as a Hidden Markov model \cite{Rabiner89atutorial, Yano2014}.This model allow us to predict how group of words are mapped to other group of words.

In the symbolic sequences of our DNA model there are forbidden words, words that were considered in \cite{Hao2007} to characterise the DNA, but in our symbolic space there are no forbidden group of words. Any 2 group of words is separated by the same small number $2L_{n}$ of nucleotides. It is the likelihood of appearance of the two separated group of words that defines the grammatical rules of our Markovian language model. 

In order to create a partition that approximately satisfies properties of a Markov system we define the correlation ($C$) and the Mutual Information Rate ($MIR$). 

Correlation is a powerful measure to stablish relationship between two variables. We measure correlation in our symbolic space by
\begin{equation}\label{eq:correlation}
C(N,\tau)=\sum_{ij} \bigg(p_{N}(i)\; p_N(i|j)_{\tau}-p_N(i)\;p_N(j)\bigg) ,
\end{equation}
where $N$ is the number of rows or columns of the partition, $p_N(i)$ is probability of points being in box $i$ and $p_N(i|j)_\tau$ is the transition probability of points going from box $i$ to box $j$ after $\tau$ iterations, $p_N(j)$ is the probability of being in box $j$. In the usual notation for the conditional probabilities, $p(i|j)_\tau$, notation adopted in this work, represents $p(j|i)_\tau$, the conditional probability of $j$ given $i$. The correlation between points and their iterations measures the degree of dependence between them. Our main hypothesis to model the DNA is to assume that there is a finite $N$ and a finite $\tau$ for which $C(N,\tau=T)\cong 0$, i.e., the DNA has mixing properties and can be described by an approximate Markov system. Notice that if $p_{N}(i)\; p_N(i|j)_{\tau}-p_N(i)\;p_N(j)=0$ for all $i$ and $j$ (strong mixing), then the partition is Markov for $\tau=T$ and generates a memoryless process, i.e.,  the points ($\delta_j$, $\gamma_j$) can be considered as random uncorrelated variables if sampled every $\tau$ iterations i.e., ($\delta_j$, $\gamma_j$), ($\delta_{j+\tau}$, $\gamma_{j+\tau}$), ($\delta_{j+2\tau}$, $\gamma_{j+2\tau}$), $\ldots$ Another way to understand the memoryless property is by noticing that if $p_{N}(j)=p_{N}(i|j)_\tau$, then the probability of being in box $j$ does not depend on which box $i$ the point was. For a mathematical demonstration see Supplementary Material. See also Supplementary Material for a detailed discussion about the relationship among correlation, mixing, and Markov chains and partitions.

The mutual information rate ($MIR$) \cite{Baptista} is given by:
\begin{equation}\label{eq:MIR1}
MIR(\delta, \gamma)=\frac{I_{s}(\delta,\gamma,N)}{T(N)},
\end{equation}
where $T(N)$ is the value such that $C(N,\tau=T$)$\rightarrow 0$. In practice $T(N)$ is the smallest value for which $\textbf{A}^{T}$ (\textbf{A} to the power of $T$) has no zero elements, a necessary condition for $C(N,\tau=T)=0$, where $\textbf{A}$ is the transitional probability matrix  with elements $p_N(i|j)_1$. $I_{s}$ is the mutual information defined as
\noindent
  
\begin{equation}\label{eq:MI}
I_{s}(\delta,\gamma)=H_{\delta}+H_{\gamma}-H_{\delta;\gamma},
\end{equation}
\noindent
where, $H_{\delta}$ = $-\sum_{i}P_{\delta}(i)$ $log(P_{\delta}(i)$), $H_{\gamma}$ = $-\sum_{j} P_{\gamma}(j)$ $log(P_{\gamma}(j)$) and $H_{\delta;\gamma}$ = $-\sum_{i,j} P_{\delta\gamma}(i,j)$ $log(P_{\delta\gamma}(i,j)$), with $P_{\delta}(i)$ representing the probability of points in column $i$ of the coordinate where $\delta$ is being plotted, $P_{\gamma}(j)$ is the probability of points in row $j$ of the coordinate where $\gamma$ is being plotted, and $P_{\delta;\gamma}(i,j)$ is the joint probability of finding points in the box $(i,j)$ formed by the overlaps of column $i$ with row $j$. Mutual information between $\delta$ and $\gamma$ measures how much $\delta$ is dependent on $\gamma$ and also the amount of information they share. As defined by Eq. (\ref{eq:MIR1}), the $MIR$ measures in average how much information per unit of ``time'' (or per symbol) the past length-$L$ symbolic sequences exchange with the future length-$L$ symbolic sequences. Then, we check for 2 criteria to search for our approximate Markov partitions:

\begin{enumerate}
\item \textbf{The mixing property}: $N=N^{*}$ and $\tau=T^{*}$ of the approximate Markov partition is found by minimising the correlation $C(N,\tau)$. This criterion is needed in order for the partition to behave as an approximate order-$\tau$ Markov partition: the correlation between words separated by $T^*$ nucleotides is close to zero. It also allows that the $MIR$ can be calculated by Eq. (\ref{eq:MIR1}).
\item \textbf{The stationary property}: $MIR(\delta, \gamma)$ obtained for $N=N^*$ is maximal and it remains invariant for any optimal value $N^*$. Invariance of $MIR$ for different $N$ reflects the fact that our Markov model is stationary with respect to the optimal values of $N$ and that the partition is generating, in the sense that information is preserved for partitions with different order and in addition an union of $4^{2}$ boxes of an order-$T_2$ partition belong to 1 box of an order-$T_1$ partition, where $T_1 < T_2$. For example, the boxes whose name are `X0.0Y' (order-4 partition) belong to the box `0.0' (order-2 partition).
\end{enumerate}

Concerning criterion 1, Fig. 2(a) shows the correlation in colour coded for different grid sizes $N$ and different $\tau$. We observe that with the increase in $\tau$ the correlation decays very rapidly. All the partitions have $\tau$ for which $C(N,\tau)\cong 0$. In other words, we can create models of the DNA considering different word lengths and different groups. From now on, however, we focus our attention into a Markovian partition that is also stationary. To that goal, we consider criterion 2.

Concerning criterion 2, as shown in Fig. 2(b), two peaks are observed for the MIR one at ${N}=4$ and another at ${N}=16$. It naturally comes from the figure that the optimal $N$ is power of 4 (excluding the peak at 14). This is not surprising since $4^{2L_n}$ is the number of different words of length $2L_n$, i.e. the number of different group of words of length $2L$. A Markov partition must have a box for each group of words. So, $N^2=4^{2L_{n}}$. From this, we conclude that $L_n=\frac{1}{2}log_{2}(N)$, so boxes have symbolic names of length $2L_n$. In these cases, notice that $T=2L_n$, the time for the correlation to decay approximately to zero. Therefore, the order $\phi$ of a partition is defined as 
\begin{equation}
\phi=2L_n=log_2(N)=T.
\end{equation}

\begin{figure}[!htb]
\begin{subfigure}{0.33\textwidth}
 \caption{}
\includegraphics[scale=0.3]{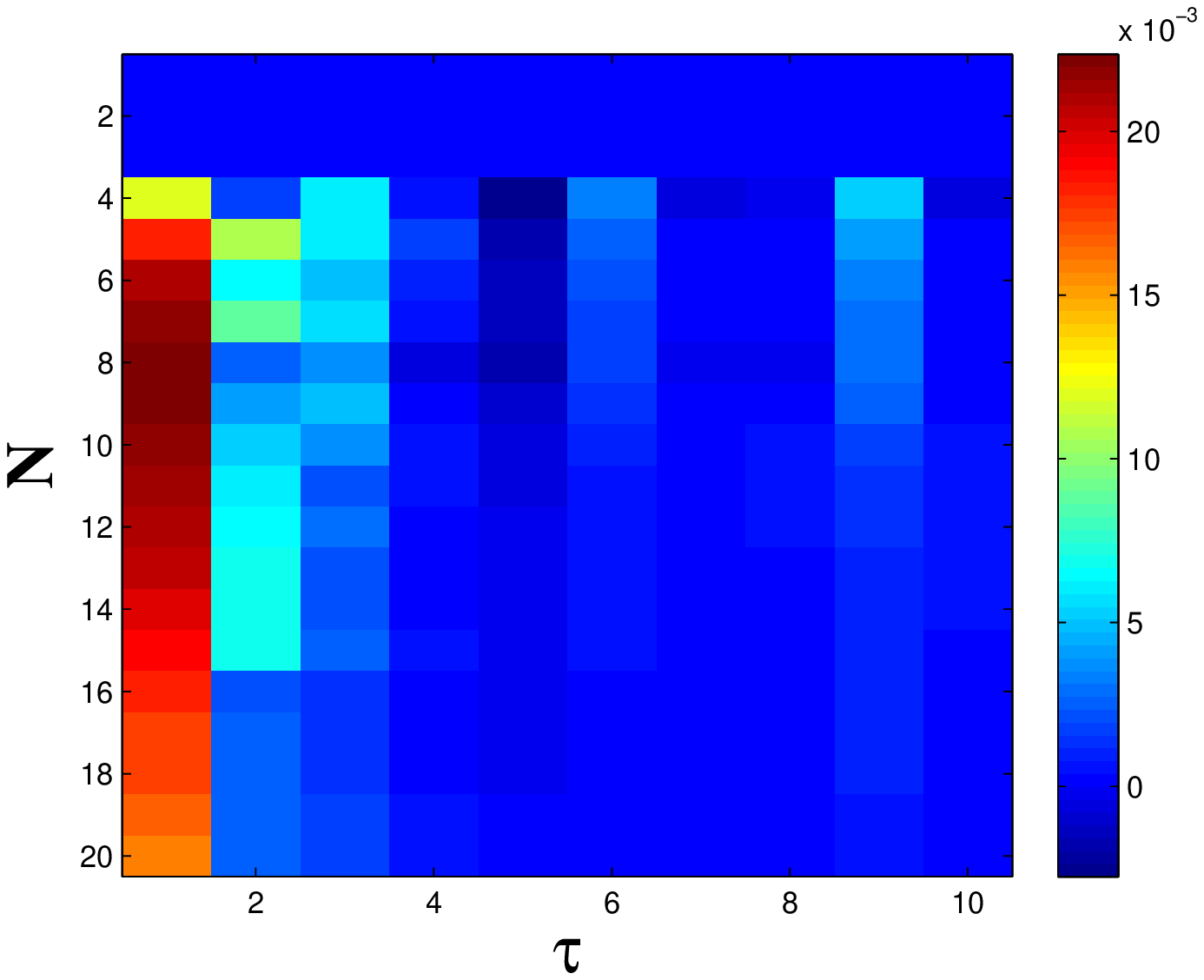}
\label{fig:correlation_all}
\end{subfigure}
\begin{subfigure}{0.33\textwidth}
 \caption{}
\includegraphics[scale=0.3]{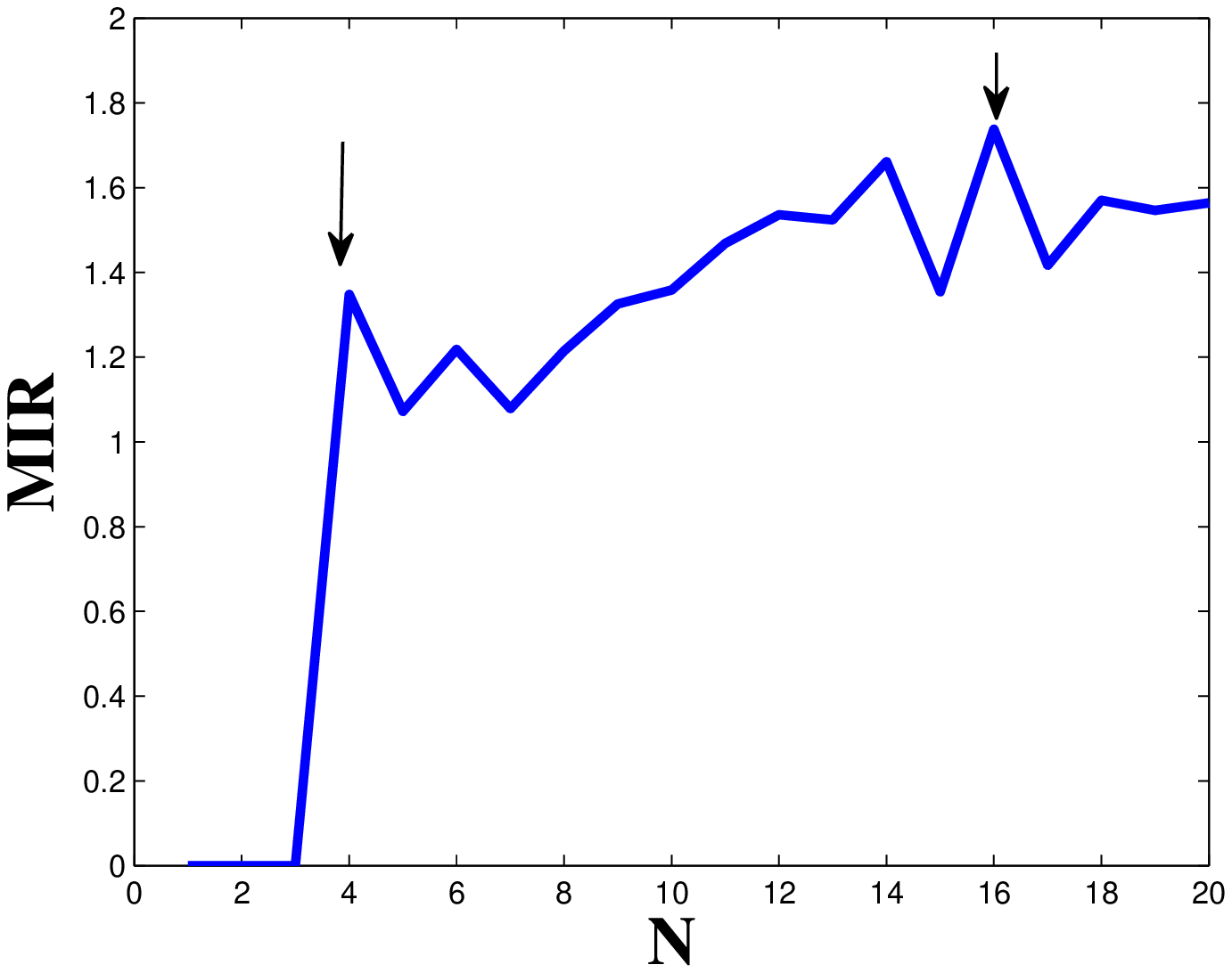}
\label{fig:MIR}
\end{subfigure}
\begin{subfigure}{0.33\textwidth}
 \caption{}
\includegraphics[scale=0.3]{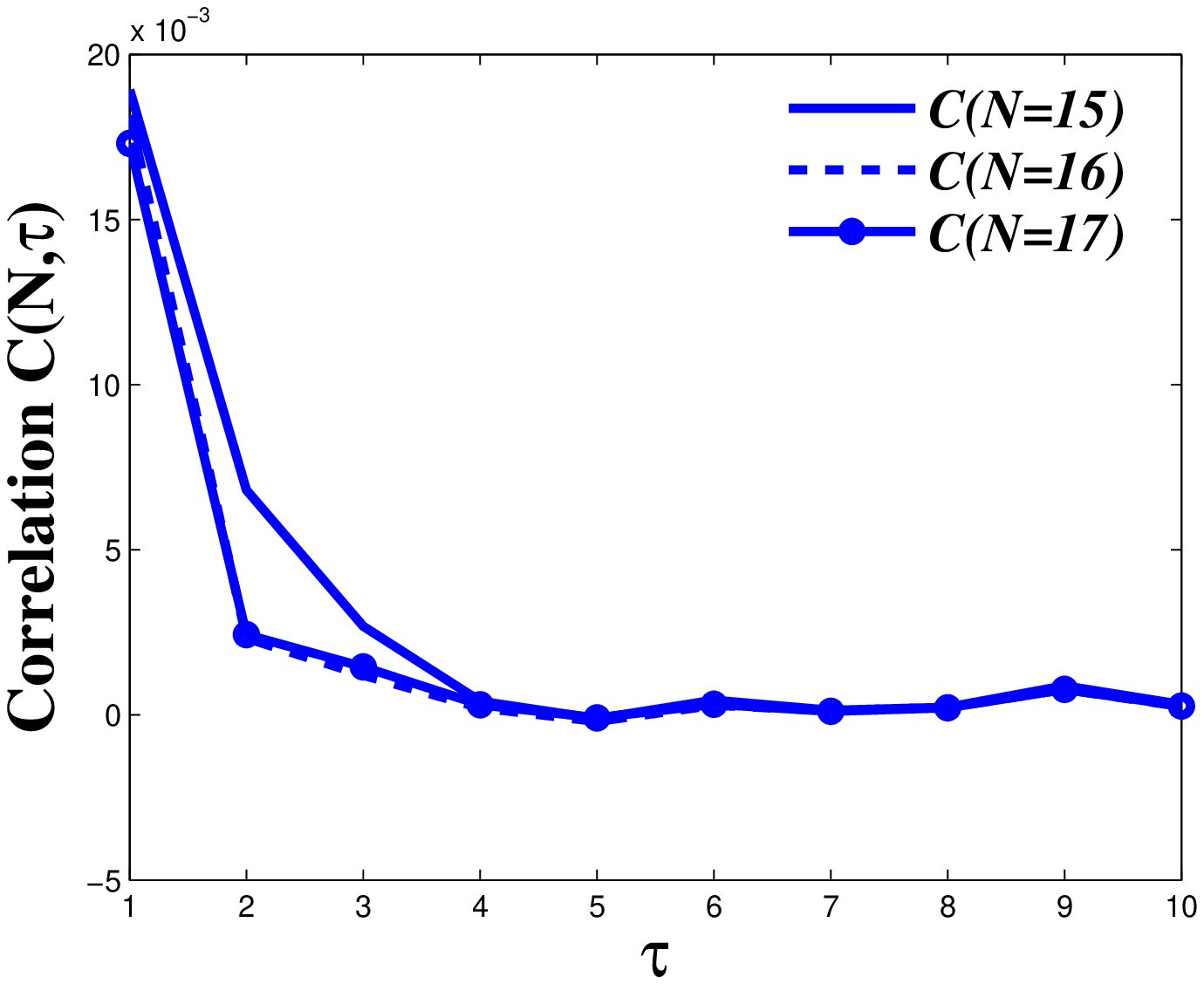}
\label{fig:correlation_without_absv2}
\end{subfigure}
\caption{Correlation and MIR for determining the resolution of the partition, which in turn provides the grouping of words. (a) Correlation for different partitions, as a function of $N$ and $\tau$ . (b) MIR as a function of $N$ for $L=4$. Maximization of MIR is attained for $N=16$. The arrow shows that maximization is possible for specific $N$. The $N$ should be a multiple of $4^n$. (c) Correlation values for $N=15$, $N=16$ and $N=17$ with respect to $\tau$. C($N,\tau$) $\cong$ 0 at around $\tau\geq4$.} \label{fig:correlation_mir}
\end{figure}

From this point on, we choose $N^*=16$, as the resolution of our optimal partition. The selected $N^*$ should also have largest correlation for $\tau=1$. In here, we want to have a Markov system for $\tau>1$, but a strongly correlated system for $\tau=1$, such that we can create network representations of the DNA where the edges represent the probability measure of points going from box $i$ to box $j$ group of words that are 1 nucleotide apart. In Fig. 2(c), we show the correlation decay for  the partitions with $N=15,16$ and $17$. In all cases, the correlation is large for $\tau=1$ and it decays fast. 

It is known that informational quantities calculated in spaces with finite resolution and with a finite number of points lead to overestimation of these quantities \cite{Steuer2002,Herzel1994}. The reason being that the probabilities calculated in a non-Markov partition would carry spurious correlations. Since our probabilistic quantities are being calculated over partitions that are approximately Markov, informational quantities obtained from them should be minimising the appearance of overestimations.

\subsection{The Markovian model of the DNA}\label{subsec_Markov_model_characters}
Each box has a name given by a symbolic sequence of length $2L_{n}$. Every point belonging to that box will have a symbolic sequence that contains the name of the box and therefore, up to $T$ iterations, we are able to have partial information about the evolution of the trajectory in the symbolic space.
Then, for $\tau=T=\phi$ iterations, the systems becomes memoryless and the transition matrix \textbf{P}$_{\tau}$ represents our Markov model of the DNA. Its main properties are:
\begin{enumerate}
\item[(i)] $C(N, \tau=\phi) \cong 0$,
\item[(ii)] $MIR(\delta, \gamma)$ is maximal and invariant over different $\phi$ orders,
\item[(iii)] $p(i|j)_{\phi} \cong p(j)$.
\end{enumerate}

Condition (i) implies that the order - $\phi$ partition provides a model that has weak mixing properties. Condition (ii) implies that we search for partitions of different orders whose content of information is invariant. Condition (iii) implies that the model also behaves as a Markov system and in addition the measure provided by the order-T partition is invariant.

In the symbolic sequences of the DNA model there are forbidden words, but there are no forbidden group words. Any 2 group of words is separated by the same small number $2L_n$ of nucleotides. It is the likelihood of appearance of the two separated group of words that defines the grammatical rules of the language model.

Because the quantities considered to model the DNA are all small and finite, computational cost can be reduced. If the Markovian property is fully verified in the DNA, or at least in a piece of it, the whole information of it can be stored in the matrix $p(i|j)_{\phi}$, which has dimensionality $4^{2\phi}$.

\section*{Results and discussion}
\section{Functional network of the genome}\label{section_functional network of the genome}

Let us define that $p(i)$ represents the probability of being in box $i$ in a partition of order $\phi=4$ (the sub index $N$ in the notation for probabilities is dropped since we set $N=16$) and $p(i|j)_1=\frac{n_\iota(i \rightarrow j)}{n_\iota(i)}$ [where $n_\iota(i \rightarrow j)$ represents the number of points in box $i$ that goes to box $j$, and $n_\iota(i)$ represents the number of points in box $i$] is a term of the transition probability matrix, representing the transition probability of going from box $i$ to box $j$, after 1 time iteration. A transition matrix represents a square matrix with positive entries such that $\sum_{j}p(i|j)_1$=1 for all i. The amount of probability measure that goes from box \textit{i} to box \textit{j} after 1 time iteration is given by $p(i)p(i|j)_1$. To create a graph representing how points in the two boxes in the partition are correlated, we consider the quantity $p(i)p(i|j)_1$. An edge connecting the node $i$ to the node $j$ is considered to exist if $p(i)p(i|j)_1\geq t^*$. There will be $N_E=4.N^2$ edges, since every node (in a partition with $N^2$ boxes) represents group of words that all contain the same sequence of length $2L_{n}= log_{2}(N)$ and shifting words in box $i$ one nucleotide (or after 1 iteration), results in another group of words that are encoded by symbols whose encoding occupy certainly $4$ other boxes (See illustrations in Fig. 1(c)). There will be $N_V=N^2$ nodes in the network, and therefore the number of edges are given by $N_E=4.N^{2}$.

\subsection{Threshold network and associated measures}\label{subsection_rescaling_matrix}

Along the lines of the work in Ref. \cite{Sinatra2010}, we apply our model to create a network representation of the DNA. In contrast to that work, our words do not necessarily have a biological meaning as the motifs considered in that work. The advantage of our approach is that being a Markov model, we can in principle predict very long sequence of symbols.  Preserving the connections responsible for most of the information of the DNA, we create a reduced network models of the DNA, which capture the most relevant features of it, and is able to identify hub groups of words, which have a similar function to the core words identified in Ref. \cite{Gerlach2013} in natural languages. 

The informational hubs of the DNA symbolic network are found by removing nodes and their edges that transfer little measure. Ignoring the smaller $p(i)p(i|j)_{1}$ values by thresholding create network representations of the DNA that contains less information about it. But selecting boxes where $p(i)p(i|j)_{1}$ are large have effect of revealing a network of very likely transitions between words, the informational hub networks. 

Removing the values from the transition matrix can change its properties entirely and it would no longer be a transition matrix which means that quantities based on probabilities such as $I_s$, $MIR$ and $C(N=16,\tau=1)$ cannot be calculated. The model would loose its Markov properties. So, to restore the properties of the transition matrix and to maintain the Markov properties of the model for a reduced network of groups of words, we have to rescale it to $\tilde{p}(i|j)_1$ so that the properties of a transition matrix are maintained, i.e., $\sum_{j}\tilde{p}(i|j)_1$=1. \\
Let the original transition matrix  for $\tau=1$ be $\textbf{A}$ with elements $\textbf{A}_{ij}=p(i|j)_1$ and the new thresholded matrix be $\tilde{\textbf{A}}$. Then we obtain $\tilde{\textbf{A}}$ by
\begin{equation}
\tilde{\textbf{A}}_{ij}=\left(\frac{\textbf{A}'_{ij}}{\sum_{j} \textbf{A}'_{ij}}\right),
\end{equation}

\noindent
where, $\textbf{A}'_{ij}=\textbf{A}_{ij}$ if $\textbf{A}_{ij}>t^*$ and $\textbf{A}'_{ij}=0$, otherwise. The rescaled network will therefore have $\tilde{N_V}$ nodes. The density $E_d$ of the edges connecting two nodes representing a symbolic transition between two group of words in the rescaled network is given by
\begin{equation}
E_d=\frac{\tilde{N_{E}}}{\tilde{N_V}},
\end{equation}
\noindent
where $\tilde{N_E}$ is the number of edges of the rescaled network, i.e., the number of times that ${p}{(i)}{p}{(i|j)}\geq t^*$. 

Total outgoing measure of the non-thresholded network can be calculated by 
\begin{equation}\label{eq:measure_network}
M=\sum_{ij}p(i)p(i|j)_1.
\end{equation}

The total measure calculated by Eq. \eqref{eq:measure_network}, but neglecting in the summation $i$ and $j$ for which $p(i)p(i|j)_{1}\leq t^*$, we denoted by $\tilde{M}$. It lies between [0, 1], where 0 means that network represents 0\% of the transitions and the measure associated to the symbolic space; 1 means that the network is based on all the observed transitions.

We must also reconstruct the invariant measure based on $\tilde{\textbf{A}}$. Given $\tilde{\textbf{A}}$ and assuming it to be a regular matrix we want to calculate the invariant measure 
\begin{equation}
\tilde{\textbf{P}}(\tilde{N}_{V})={\tilde{p}(i), i=1,\ldots, \tilde{N}_V},
\end{equation}
such that $\sum_{i=1}^{\tilde{N}_V} \tilde{p}(i)$=1. Here we assume that the nodes of the thresholded network represent group of words where probability of appearance is provided by a Markov memoryless process. This implies that if $\tilde{\textbf{P}_{0}}$ represents a vector of initial random values with $\tilde{N}_V$ entries forming a vector, then $\tilde{\textbf{P}}$ with elements $\tilde{p}(i)$ is calculated by
\begin{equation}
\lim_{n\rightarrow\infty}\tilde{\textbf{P}}^{n}=\tilde{
\textbf{P}}_{0}\;\tilde{\textbf{A}}^{n},
\end{equation}

In order to characterise a system whose probabilities and transition probabilities are given by $\tilde{p}(i)$ and $\tilde{p}(i|j)_1$ respectively, we calculate the $MIR$, which measures exchange of information per symbol between two group of words of length $2L$ shifted 1 nucleotide apart defined as
\begin{equation}
\tilde{MIR}(\tilde{N}_V,t^*)=\frac{I_s(\tilde{N}_V,t^*)}{\tilde{T}},
\end{equation} 
where $I_s$ is defined as
\begin{equation}\label{eq:threshold_mi}
I_s(\tilde{N}_V,t^*)=\sum_{i,j}\;\tilde{p}(i)\tilde{\textbf{A}}_{ij}\;log  \left(\frac{\tilde{p}(i)\tilde{\textbf{A}}_{ij}}{\tilde{p}(i)\tilde{p}(j)}\right).
\end{equation} 
Notice that $\tilde{p}(i)\tilde{A}_{ij}$ is a joint entropy. The average rate of information contained in all groups of words is calculated by Shannon's entropy rate
\begin{equation}
S(\tilde{N}_V,t^*)=\frac{S_{n}(\tilde{N}_V,t^*)}{\tilde{T}},
\end{equation}
where $S_n$ is defined as
\begin{equation}
S_{n}(\tilde{N}_V,t^*)=\sum_{i}\Big(\tilde{p}(i)\;log\;\frac{1}{\tilde{p}(i)}\Big).
\end{equation}
$\tilde{T}$ is the time for which $\tilde{\textbf{A}}^{\tilde{T}}$ has only non-null elements. If the network is not connected (it can be decomposed in 2 or more sub-networks), then $\tilde{T}$ is the average of all $\tilde{T}$ for each sub-network.
To determine how the mutual information changes when we remove the nodes, we calculate the mutual information $I_s$ between two group of words per node $\tilde{N_V}$ for each threshold $t^*$ 
\begin{equation}
\sigma_I=\frac{I_s(\tilde{N}_V,t^*)}{\tilde{N}_V}.
\end{equation}
\noindent
Similarly, we calculate the Shannon entropy per node also, defined by
\begin{equation}
\sigma_S=\frac{S_{n}(\tilde{N}_V,t^*)}{\tilde{N}_V}.
\end{equation}

\begin{figure}[!h]
\begin{subfigure}{0.4\textwidth}
\caption{}
\includegraphics[scale=0.46]{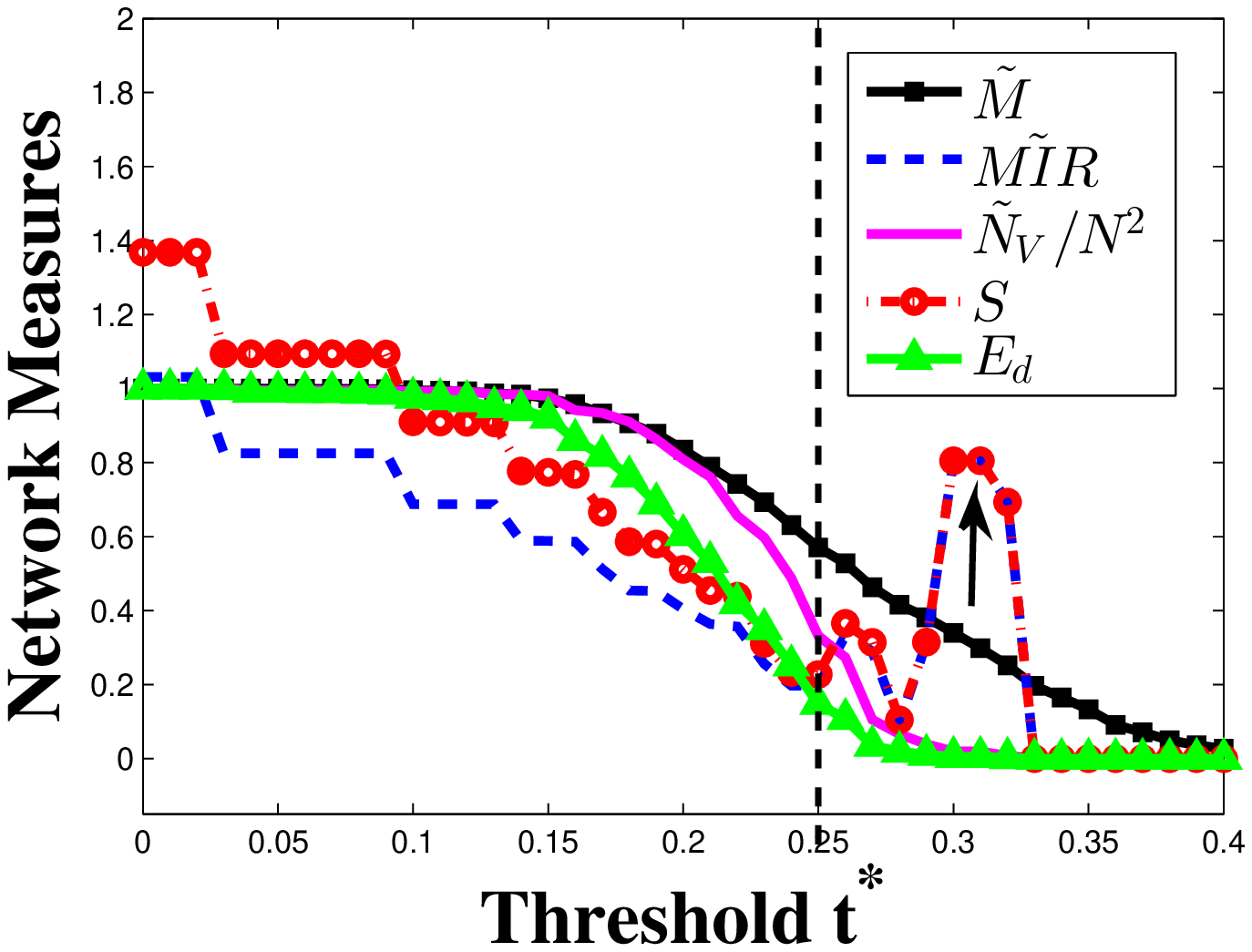} 
\label{fig:threshold}
\end{subfigure}
\hspace*{15 mm} 
\begin{subfigure}{0.4\textwidth}
\caption{}
\includegraphics[scale=0.45]{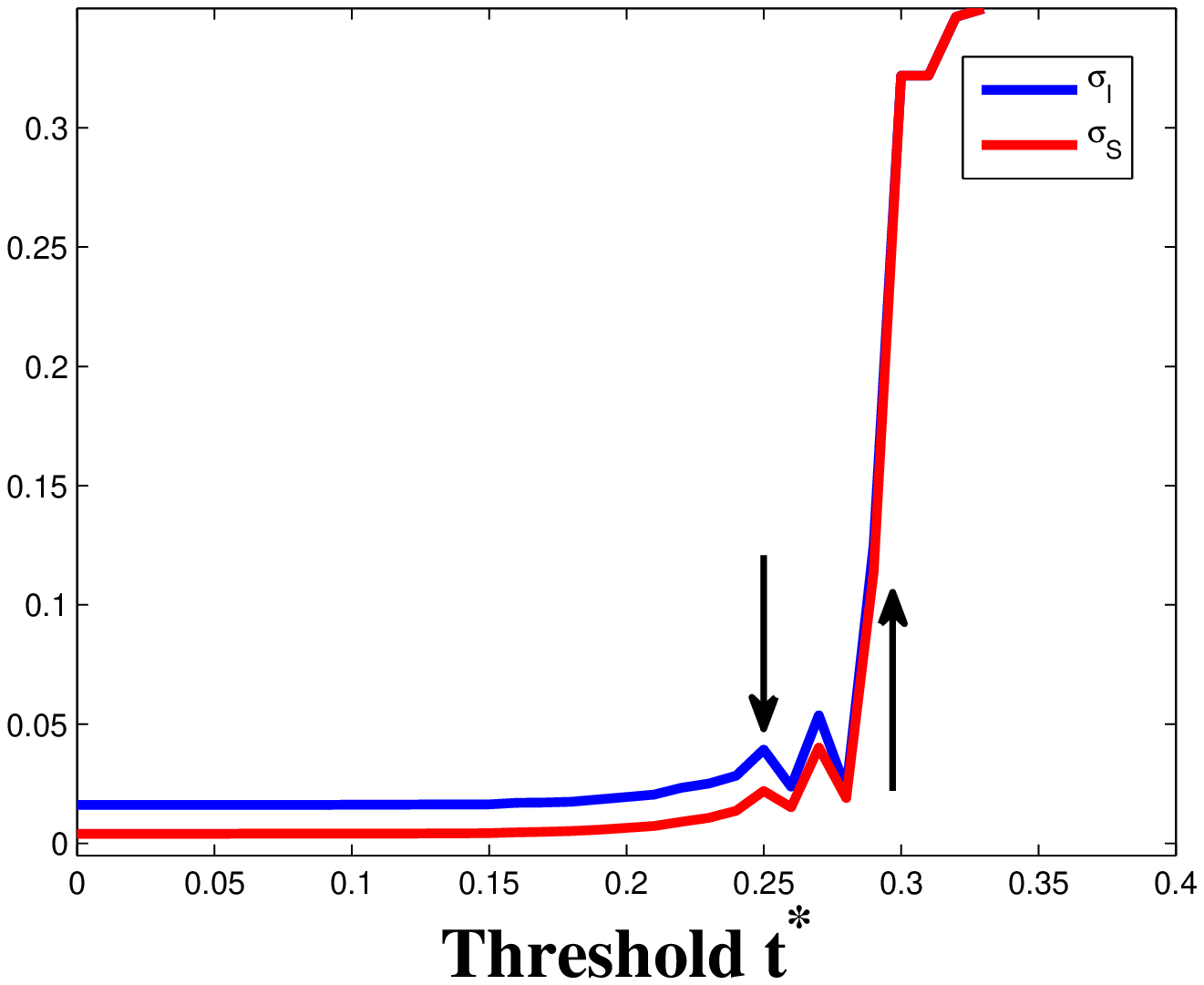}
 \label{fig:MI_S}
\end{subfigure}
\caption{Characterisation of a network representation of DNA based on information theoretical quantities as a function of the threshold $t^*$. (a) $\tilde{M}$ is the total probabilistic measure of the thresholded network (black line and square online), $MIR$ is the mutual information rate (blue dashed line online), pink line online represents the relative number of nodes $\frac{\tilde{N}_V}{N^2}$, $S$ is the Shannon entropy rate (red line and circles online) and $E_{d}$ is the density of edges (green line and triangle online). The dashed line represents the $t^*=0.25$. (b) The curves show how information grows with the decrease in the number of nodes but retaining only nodes with high information content. $\sigma_I$ is mutual information per node and $\sigma_S$ is Shannon entropy per node. The arrows represent the unique threshold points where the information changes.}
\label{fig:network}
\end{figure}

Fig. 3(a) shows different informational quantities and their behaviour as we change the threshold $t^*$. The knock-out of only 1 group of words for $t^{*}=0.03$ leads to an approximate 20\% of reduction in the information produced $(S)$ and exchanged by the network. When compared with the decrease in the number of nodes at this $t^*$, this number changes only by 1 but the change in $\tilde{MIR}$ and $S$ shows evident loss of information. For $t^* \in$ [0.1, 0.13] $\tilde{MIR}$ and $S$ remain constant and decreases for $t^* \cong$ 0.14. This pattern of $\tilde{MIR}$ and $S$ of being constant and then decreasing goes until $t^* =$ 0.24. The picture shows 3 changing steps for $\tilde{MIR}$ and $S$ for $t^*\leq0.18$. Each step represents a single node in the thresholded network being eliminated. The nodes lost respectively, for these three steps represent the `CTAG', `CCTG' and `TTAG' group of words. These words belong to the well defined Group-I and Group-II tetramer class of nucleotides \cite{Gabriel1994}. Studies have shown that these group of tetramers are actively involved in base mismatch repair in \textit{E. coli} and are known as Very Short Patch (VSP) \cite{Bhagwat1992}. The word `CTAG' is a well-known palindromic sequence which is rarely present in protein-coding region but is abundantly present in genes coding for structural RNAs \cite{Blattner1997}. Almost $\frac{2}{5}$ of the information of the DNA is contained in symbolic sequences formed by these 3 length-4 words. A smooth decay in $\tilde{M}$ and $\frac{\tilde{N}_{V}}{N^{2}}$ is observed for $t^*\in$ [0.12, 0.20] but suddenly an abrupt change is noticed in $\tilde{M}$ and $\frac{\tilde{N}_{V}}{N^{2}}$ for the interval $t^*\in$ [0.2, 0.25]. The number of nodes of the thresholded network change from 256 at $t^*=0$ to 85 at $t^*=0.25$ after this abrupt decay. Although the percentage of nodes remaining is just $\sim$ 33\% of all possible nodes, the thresholded network's nodes contain words which appear in most of the genes and $\tilde{M}$, $\tilde{MIR}$ and $S$ are moderately high. Moving further to $t^*$ $\in$ [0.25, 0.3], $\tilde{MIR}$ and $S$ increases for $t^*$ $\in$ [0.25, 0.27] and then decays abruptly for $t^*$ $\in$ [0.27, 0.3]. The number of nodes of the rescaled network for $t^*$ $\in$ [0.25, 0.3] change from 85 at $t^{*}=0.25$ to 5 at $t^{*}=0.30$. For $t^*\geq$ 0.3, the number of nodes remains constant, and not only $S = \tilde{MIR}$ but both quantities are very high. This shows that the minimum number of nodes that remain even after thresholding the data is 5. 

The fact that $S = \tilde{MIR}$ means that all the measure of a box is mapped to another unique box, and consequently in Eq. \eqref{eq:threshold_mi}, $\tilde{p}(i)\tilde{\textbf{A}}_{ij} = \tilde{p}(j)$. Group of words map uniquely to another group and every node of the network has a degree 1.
To make this argument rigorous, if $\tilde{p}(i)\tilde{\textbf{A}}_{ij}=\tilde{p}(i)$, it means that $\tilde{\textbf{A}}_{ij}=1$ for that $i$. But $\sum_{j}(\tilde{\textbf{A}}_{ij})=1$ which means that $\tilde{\textbf{A}}_{ij} = 0\mbox{ for i} \neq j $. So, we can write Eq. \eqref{eq:threshold_mi} as 
\begin{equation}\label{eq:new_mi_i_eq_j}
\tilde{I_s}(\tilde{N}_V,t^*)=\sum_{i}\sum_{j=i}\;\tilde{p}(i)\;\log  \left(\frac{\tilde{p}(i)}{\tilde{p}(i)\tilde{p}(j)}\right).
\end{equation}
and since $j=i$ in the summation of Eq. \eqref{eq:new_mi_i_eq_j}, Eq. \eqref{eq:threshold_mi} can be written as
\begin{equation}
\tilde{I_s}(\tilde{N}_V,t^*)=\sum_{i}\;\tilde{p}(i)\;\log  \left(\frac{1}{\tilde{p}(i)}\right).
\end{equation}
Therefore, 
\begin{equation}
\tilde{I_s}(\tilde{N}_V,t^*)=S_{H}(\tilde{N}_V,t^*).
\end{equation}

In Fig. 3(b), we show how $\tilde{MIR}$ and $S$ increase per $\tilde{N_V}$. Both curves are roughly constant for $t^*\in$ [0, 0.18]. For larger $t^*$ values these quantities start to fluctuate. A small peak is observed at $t^*=0.25$ for both quantities while few fluctuations happens for the range of $t^* \in$ [0.25, 0.29] but then these quantities abruptly increase at $t^*=0.30$. This is not surprise, since $\tilde{I_s}$ maximises when $\tilde{p}(i)\tilde{\textbf{A}}_{ij} = \tilde{p}(j)$. These results suggest that the networks obtained for the interval $t^* \in$ [0.25,0.30] represent group of words that exchange high amounts of information (relative per remaining nodes). Therefore, if a word is found in the DNA that belongs to one of the groups of these rescaled networks the predictability of the possible iterated word is very high.

\begin{figure}[!h]
\begin{subfigure}{0.4\textwidth}
\caption{}
\includegraphics[scale=0.4]{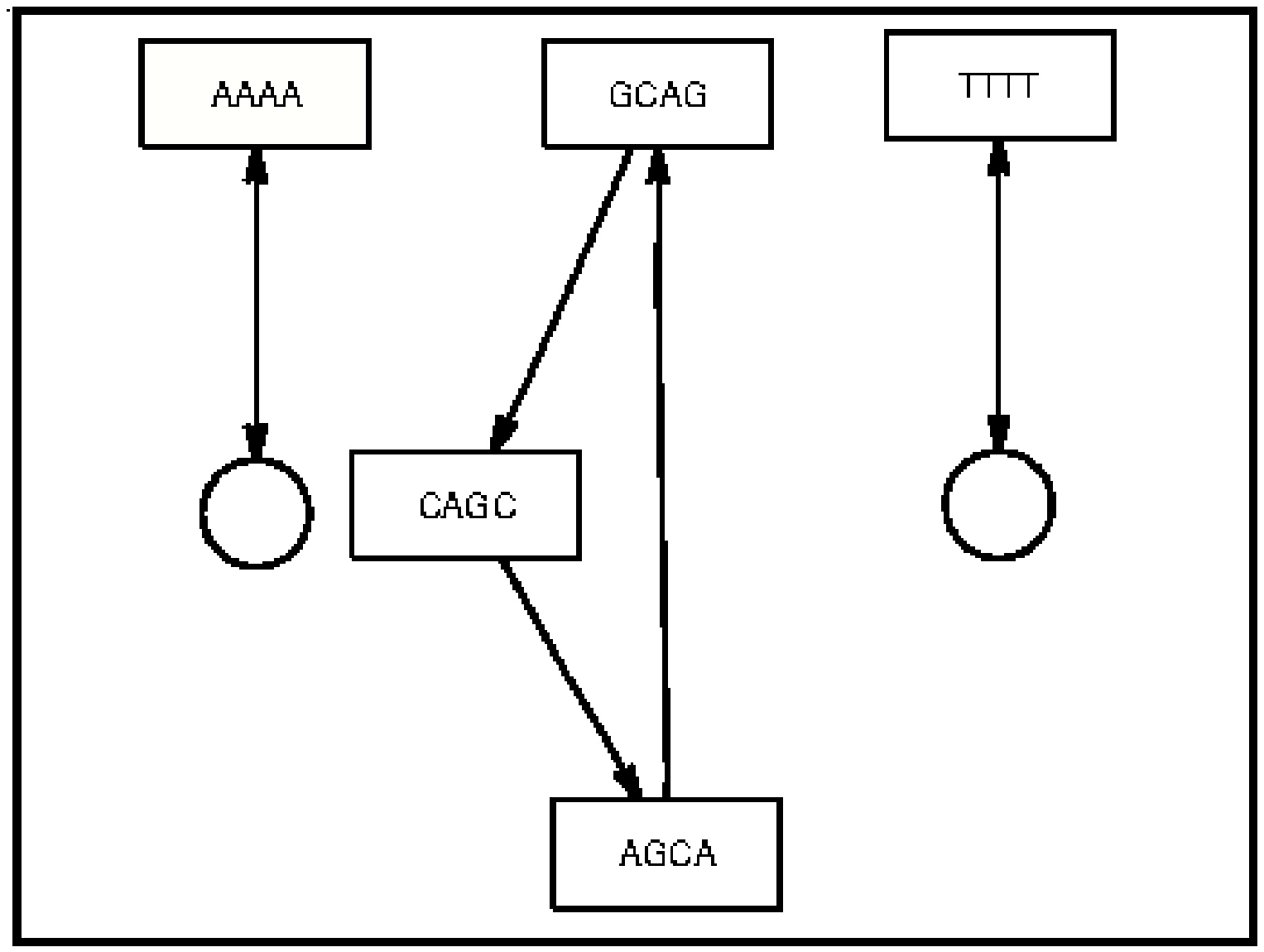} 
\label{fig:5_nodes}
\end{subfigure}
\hspace*{10 mm} 
\begin{subfigure}{0.4\textwidth}
\caption{}
\includegraphics[scale=0.4]{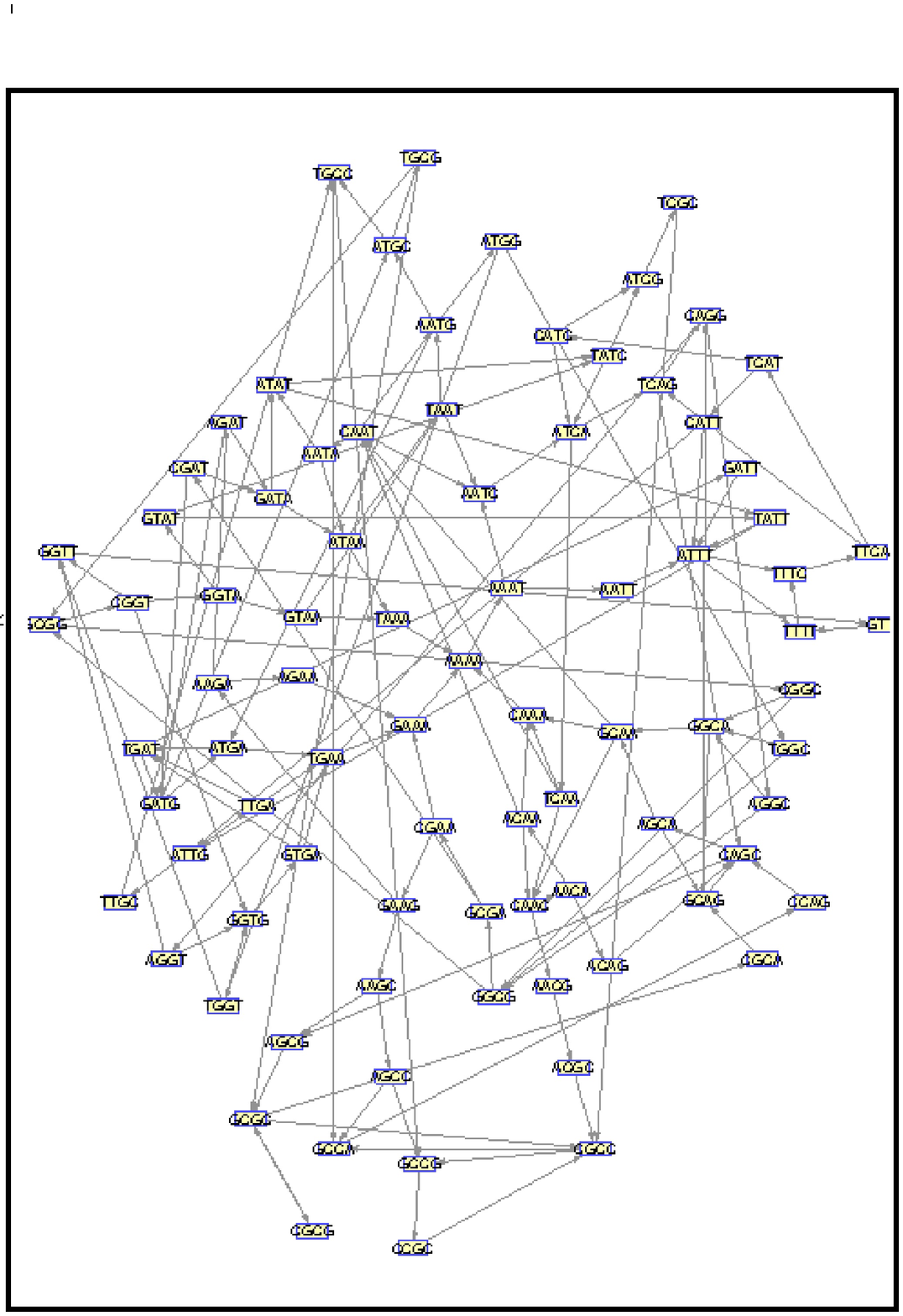}
\label{fig:46_node}
\end{subfigure}
\caption{(a) Network with 5 nodes at t$^{*}$=0.3. In the center, the words GCAC, CAGC, AGCA are connected in a cycle of length 3. The rectangular boxes represent the nodes and the circles are used to represent a self-loop. (b) Network with 85 nodes at t$^{*}$=0.25.}
\label{fig:rescaled_network}
\end{figure}

In Fig. \ref{fig:rescaled_network}, we show 2 examples of the rescaled networks obtained.	The smallest network obtained for $t^*=0.30$ with 5 nodes is shown in Fig. 4(a). The sequence generated from the cycle of length 3 connecting three nodes is ``GCAGCA''. This sequence is formed by a likely transition of words of 4 symbols that contain high amounts of information. Therefore, one should expect the appearance of this sequence in the DNA of \textit{E. coli}. One should also expect these transitions between the length-4 words. From Fig. 3(a), one sees that this network at $t^*=0.3$ has values of $MIR$ and $S$ comparable to the larger network at 85 nodes (at $t^*=0.25$). From Fig. 3(b), it is clear that the information content per node of this network is one of the highest. 

In Fig. 4(b) we show a network of 85 nodes. This network contains only $\sim$ 33\% of all possible nodes, but surprisingly at $t^*=0.25$, $\tilde{M}$ is about half the total measure of the full network, which means that these words are very likely in the genome. Remarkably, this network is the first smallest network that can generate large sequence of words with any periodicity, since it has cycles that are connected and that form words that appear with periodicity of 1, 2 and 3. This network reminds us of the Sharkovskii's theorem which states that if a dynamical system with some properties has an equilibrium point, a period 2 and a period 3 orbit then it must have periodic points for every other period. This network is very special. From Fig. 3(a) , we conclude that adding just few more edges in this network (decreasing from $t^*$ from 0.25), we would restore almost all the information content of the \textit{E. coli} DNA. If only a few more nodes are removed (increasing $t^*$ from 0.25) we see that information ($I_s$ and $S$) abruptly decays. This network therefore should represent the structural ``skeleton'' of the \textit{E. coli}, i.e., the sequences generated by it should be fundamental to the \textit{E. coli}. The analogy of the network structure possessing a skeleton property reflects the idea that a skeleton provides the structural stability of a configuration. Removing a piece of it leads to an unstable configuration. Introducing more structure, the configuration becomes stronger and more stable. To understand the importance of networks shown in Fig. \ref{fig:rescaled_network}, we created a genomic sequence from the transitions in the group of words depicted by the networks and matched them with the known genes of \textit{E. coli}. Some of the genes which contained sequence from both the networks shown in Fig. \ref{fig:rescaled_network} are nrfA, flgF, mnmC , cysM, xseA, hscB. These genes are known to be involved in structural foundation of the organisms by either being a structural genes or producing proteins for structural stability. They are also involved in DNA-binding and promoting stress-induced mutagenesis. \cite{Eaves1998, Jones1990, AlMamun2012}. We also found iscS, iscU and iscR which help in DNA-binding and scaffold protein for iron-sulphur cluster assembly. These 3 genes work along with hscB and are involved in some pathways like alanine biosynthesis III, molybdenum cofactor biosynthesis and thiazole biosynthesis I \cite{Keseler2013}.

\section{Validation of the model and predictability of group of words for genes}\label{section_validation_of_the_model}
To validate the model and study its predictability we transform $\tilde{\textbf{A}}({t^*})$ from the whole genome to an adjacency matrix $\textbf{G}^{E.coli}(t^*)$, with elements ${G}_{ij}^{E.coli}(t^*)=1$ if the group of words in box $i$ iterate to box $j$ or ${G}_{ij}^{E.coli}(t^*)=0$ if there are no transitions from box $i$ to box $j$. Then, we evaluate the efficiency of this model at different levels of $t^*$, by studying its ability to predict genes $g_i$ at $t^*$ levels. We create a symbolic sequence for each $g_i$, similar to the method we followed for genomic sequence (N=16, L=4), and obtain an adjacency matrix $\textbf{G}(g_i,t^*)$ from the transition matrix $\textbf{B}(g_i,t^*)$ for gene $i$. ${G_{ij}}(g_i,t^*)=1$ if the group of words where points are in a box i are iterated to box j (so $B_{ij}(g_{i},t^*)>$0, and $G_{ij}(g_{i},t^*)=0$, if $i = j$ or $B(g_{i},t^*) = 0$, otherwise). We then compare $\textbf{G}(g_{i},t^*)$ with $\textbf{G}^{E.coli}(t^*)$, the adjacency matrix for the genome of $E.coli$ defined at different threshold $t^*$ and see how efficiently we can predict the existence of group of words and their transitions in each gene. For this purpose, we calculate parameters for the True Positive Rate (TPR) or sensitivity $S_{n}$, and False Positive Rate (FPR) or specificity $S_{p}$ of the model. These parameters are defined as 

\begin{equation}\label{eq:specitivity}
S_n=\frac{T_P}{T_P+F_N},
\end{equation}
\begin{equation}\label{eq:sensitivity}
S_p=\frac{T_N}{T_N+F_P},
\end{equation}

\noindent
where TP is the number of transitions between 2 groups of words correctly predicted, therefore, $TP$ for gene $g_i$ is defined as $TP^{E.coli}(g_{i}, t^*)=\sum_{ij}^{1,1}(G_{ij}^{E.coli}(t^*)-G_{ij}(g_{i},t^*))$, where we only take into consideration all the $i$ and $j$ values of $G_{ij}^{E.coli}(t^*)$ and $G_{ij}(g_{i},t^*)$ which are equal to 1. The symbol $\sum_{ij}^{1,1}$ represents a summation that is only carried out when the variables inside the argument are equal to the super index. FN is the number of words that were wrongly predicted, $FN^{E.coli}(g_{i}, t^*)=\sum_{ij}^{1,0}(G_{ij}^{E.coli}(t^*)-G_{ij}(g_{i},t^*))$, this can happen only when $G_{ij}^{E.coli}(t^*) = 1$ and $G_{ij}(g_{i},t^*)=0$, meaning that a transition from the group of words in box $i$ are mapped to box $j$ are not present but have been wrongly predicted by the model. $TN^{E.coli}(g_{i}, t^*)=\sum_{ij}^{0,0}(G_{ij}^{E.coli}(t^*)-G_{ij}(g_{i},t^*))$, in this case we consider all the values of $G_{ij}^{E.coli}(t^*)$ and $G_{ij}(g_{i},t^*)$ that are equal to zero, meaning that the a transition from the group of words in box $i$ are mapped to box $j$ do not exist and the model also does not predicts them. $FP^{E.coli}(g_{i}, t^*)=\sum_{ij}^{0,1}(G_{ij}^{E.coli}(t^*)-G_{ij}(g_{i},t^*))$, happens when $G_{ij}^{E.coli}(t^*)=0$ but $G_{ij}(g_{i},t^*)=1$. 

\begin{figure}[!h]
\centering
\includegraphics[scale=0.50]{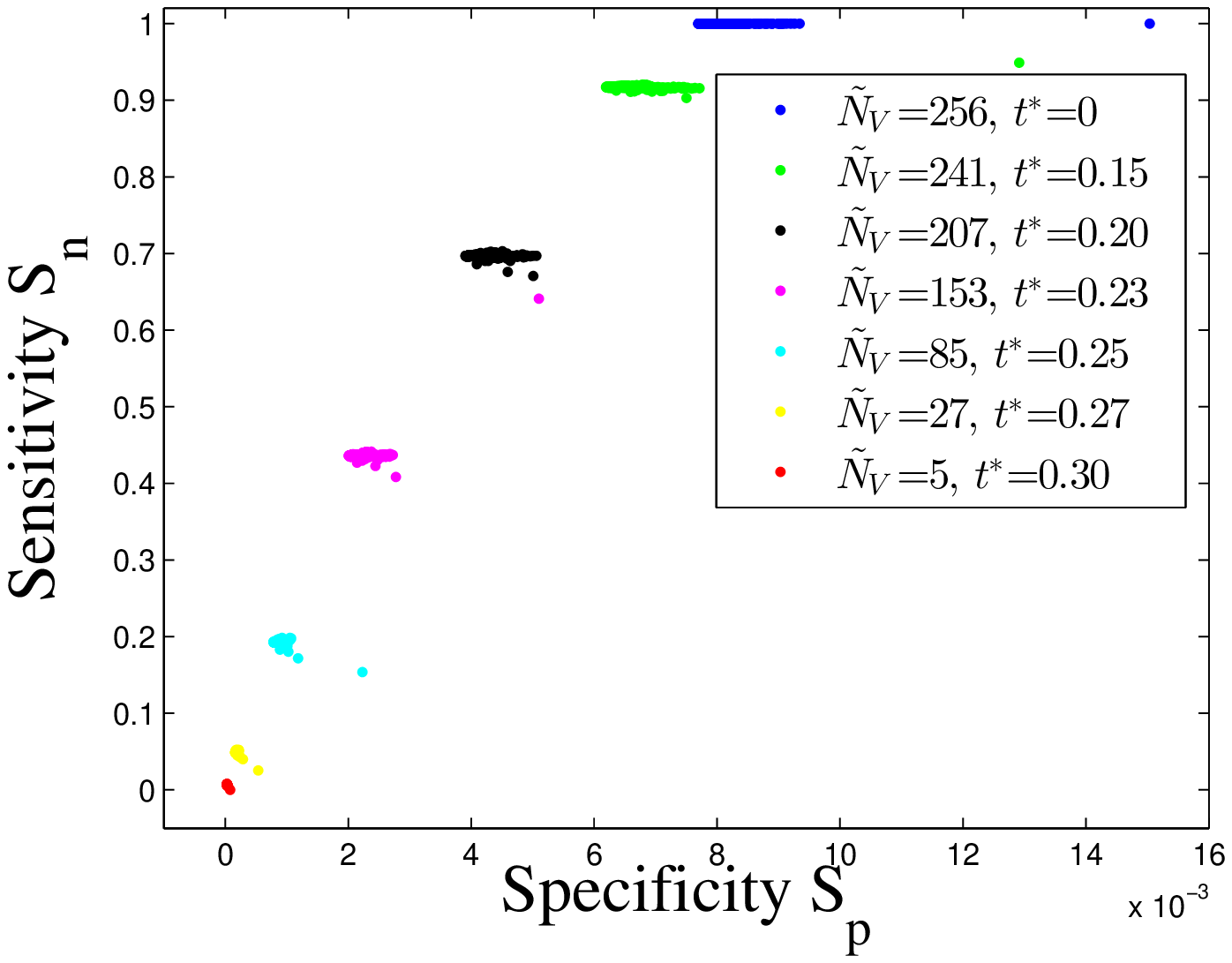}
\caption{Specificity $S_p$ shown on horizontal axis and Sensitivity $S_n$ shown on vertical axis plot for \textit{Escherchia coli} with different levels of threshold $t^*$.}
\label{fig:roc_ecoli}
\end{figure}

Figure \ref{fig:roc_ecoli} shows the $S_n^{E.coli}(t^*)$ vs $S_p^{E.coli}(t^*)$ plot for the different thresholds $t^*$ considered to construct $\textbf{G}^{E.coli}(t^*)$ and $\textbf{G}(g_{i},t^*)$. The results obtained for each $t^*$ network has been represented with different colors and symbols. The results obtained for the network composed of 256 nodes ($t^*=0$) is shown with blue points. The $S_n^{E.coli}(t^*)$ value for this network is $\simeq$ 1 and $S_p^{E.coli}(t^*) \cong 0$, meaning  that this network correctly predicts the transition of groups of words for genes. Having the TPR or $S_n$ constant for all genes at a given threshold shows that our model can predict every gene with similar accuracy. Regardless of the fact that the model was constructed from the whole genome data, we could still predict transition of groups of words in each gene. Notice that how words are mapped to words is not relevant in our model, but how group of words are mapped to other group of words. This is a consequence of the Markovian characteristics of the model that the property of the ``whole" reflects also the property of the ``parts".

\noindent
In Supplementary Material we show how our model can be used to detect similarity between genes of different organisms and shows that our model predicts better the appearance of group of words then standard probabilistic  methods that predicts the appearance of particular words.

\section{Data collection}\label{section_data_collection}
All the genomic sequences for organisms were download from NCBI(http://www.ncbi.nlm.nih.gov/). The NCBI Reference Sequence id for each organisms are as follows:[\textit{E. coli}: GenBank:\texttt{NC\char`_000913.3}, \textit{Shigella dysenteriae}: GenBank:\texttt{NC\char`_007606}, \textit{Rhodococcus fascians}: GenBank:\texttt{NC\char`_021080.1} and \textit{Saccharomyces cerevisiae}: GenBank:\texttt{NC\char`_001133.9}].

\section{Conclusion}\label{section_conclusion}
This work proposes a Markovian language model for the DNA. We encode nucleotide sequences into symbolic sequences of finite length, regarded as words, and then encode these words into points belonging to a 2D symbolic space of the DNA; from which we establish the functional connectivity between any two regions in this symbolic space, representing two groups of similar words, i.e., the likelihood of having a word belonging to a group being followed after some nucleotides into another word that belongs to another group of words. 

We construct a Markov network representation of the DNA, the nodes representing group of words and their probabilities and the edges their transition probabilities, such that the statistics of the transition between group of words is approximately memoryless.  Our model allows for a reduction in the complexity of the DNA. For the \textit{E. coli}, we showed that a network of only 85 nodes (representing 85 group of words) contains most of the information of its DNA, composed of more than 1 billion nucleotides. On the other hand, a network with all but 3 words looses almost $\frac{2}{5}$ of its information content. We have also shown that our model can be used as a similarity measure to detect similar symbolic genes in different organisms (See Supplementary Materials). The results demonstrated that genes with not entirely known function were similar to genes with well established and known function as in case of \textit{E. coli} and \textit{Saccharomyces cerevisiae}. 

We have shown that for the DNA, an approximate Markov partition can be constructed assuming equal-sized cells. Among all possible sizes, we have shown that our approximate Markov partition exists when there are $4^L$ cells. Our approach can be extended to other systems as long as a Markov partition is obtained. To construct Markov partitions for systems presenting time-varying delays or partially accessible information one could consider the works in Refs. \cite{WeiPQJ15, WeiQKW13}, or the work in Ref. \cite{Bolt2008} for stochastic systems, or the work in Ref. \cite{Hirata2013} for low-dimensional dynamical systems.

This model can also be used to calculating the statistics of recurrence of group of words in DNA (See Supplementary Materials). Studying the recurrence of these group of words can be used in determining the stochastic and the deterministic nature of the DNA which contributes to the genes, coding and non-coding regions of the DNA, and provide an explanation for the evolution of the DNA from simple organisms to the Human genome in terms of how the recurrence of the DNA has become more or less memoryless. Our model allows to deduce analytical expressions for the probability density of returns of group of words. The more Markovian a piece of the DNA is, the more accurate are our analytically obtained density. In a publication to be submitted elsewhere, we will demonstrate that the coding part of the DNA is more Markovian (random) than the non-coding, and the sequences appearing on it can have their short and long-term behaviour well predicted by our model.      

\section*{Competing interests}
  The authors declare that they have no competing interests.

\section*{Author's contributions}
    SS has constructed all the matlab codes used for this work, done all the data mining and analysis, made the conceptual links between the results of this work with the genes and function of the genes, and has independently derived some mathematical expressions. SS and MSB have together developed all mathematical and conceptual framework behind the construction of this Markovian model, have done other mathematical derivations, and written the manuscript. MSB has acted as the PhD supervisor of SS, having initially proposed the topic of study and the theoretical approaches to be considered.
\section*{Acknowledgements}
 SS and MSB acknowledge the Engineering and Physical Sciences Research Council (EPSRC), grant Ref. EP/I032608/1.
\newpage

\end{document}